# Astronomische Nachrichten – Broadband photometry of asteroid 6478 Gault: activity and morphology

S. Borysenko[1], A. Baransky[2], A. Simon[3], and V. Vasylenko[3]

[1] Main Astronomical Observatory of NAS of Ukraine, Akademika Zabolotnoho 27, Kyiv 03143, Ukraine
[2] Astronomical Observatory of Taras Shevchenko National University of Kyiv, Observatorna Str. 3, Kyiv 04053, Ukraine,
[3] Taras Shevchenko National University of Kyiv, Glushkova Ave. 4, Kyiv 03022, Ukraine

We present the results of short observational program of asteroid 6478 Gault with using of *V* and *R* Johnson filters realized at the Kyiv Comet Station during January – April 2019. Color indices and distribution of brightness in the comet-like tail were calculated. We made a comparative analysis of circumstances of the asteroid 6478 Gault activity with others morphologically similar active Main Belt asteroids.

## 1 Introduction

Active Asteroids (AAs) are small bodies with orbits typical of asteroids that exhibit cometary activity (Jewitt 2012). The first oficially registered discovering of activity in a main-belt asteroid was more than 20 years ago for 7968 Elst – Pizzarro object well known as comet 133P/Elst – Pizzarro. For today about two dozen active main-belt objects were registered. Some of them repeat their activity (appearing of small coma or tail) occasionally as a result of ice sublimation. Usually, we call them the main-belt comets (MBCs). But some part of the objects no more signs of activity (like 596 Scheila) and still have a starlike asteroidal image. Temporally activity of such objects cannot be explained by ice sublimation. Usually, the events of recurrent activity are much less likely to be caused by the sublimation of volatiles. Some of these phenomena can be explained by collision with another asteroidal body (as in the case of 596 Scheila). But for so small size objects probability of such collisions still very low (Farinella & Davis 1992; Menley et al. 1998; Chandler et al. 2018).

In January 2019, it was found that one of Phocaea asteroid from the inner regions of the asteroid belt, 6478 Gault, shows cometary activity, and that it has a long thin tail. Detailed historical analysis of Gault behavior showed a obvious activity of the object in past (Smith et al. 2019; Chandler et al. 2019; Marsset et al. 2019; Moreno et al. 2019).

Orbital parameters of asteroid Gault are close to orbits of some active asteroids (such as 354P/LINEAR), but its orbit has more high inclination ($i = 22.8$) (Table 2). Usually, to classify objects by orbital location the Tisserand parameter is used. While all the orbital parameters of an object orbiting the Sun during the close encounter with another massive body (e.g. Jupiter) can be changed, the value of a function of these parameters, called Tisserand's parameter ($T_J$) is approximately conserved, making it possible to recognize the orbit after the encounter. For objects in the main asteroid belt $T_J$ = 3.

Phocaea asteroids have orbits with semi-major axes between 2.2 AU and 2.5 AU, an eccentricity greater than 0.1, and inclination between 18 and 32 degrees. The Phocaea asteroids are located close to Hungaria group (Elkins-Tanton 2010), but the division between the two groups is real and caused by the 4:1 resonance with Jupiter. The family has an estimated age of 2.2 billion years and derives its name from its most massive member, 25 Phocaea, which is about 78 km in diameter. The Phocaea group is composed mainly of S-type asteroids with a spectral type that is indicative of a stony mineralogical composition (Carvano et al. 2001; Knezevič & Milani 2003; Carruba 2009; Nesvorny 2014).

Presumably, the asteroid 6478 Gault began to break up as its spin accelerated due to the YORP effect (Yarkovsky – O'Keefe – Radzievskii – Paddack effect), and its rotation speed approached two hours – near the limit of stability for an asteroid. The YORP effect changes the rotation state of a small astronomical body – that is, the body's spin rate and the obliquity of its pole(s) – due to the scattering of solar radiation of its surface and the emission of its own thermal radiation (O'Keefe 1976; Paddack 1969; Radzievskii 1954; Rubincam 2000). As results of YORP effect influence, the ejected matter can create more than one tails. Thus, we can have a new type of objects – objects with persistent dust activity caused by rotation (Chandler et al. 2019; Jewitt et al. 2019).

---

⋆ Corresponding author: borisenk@mao.kiev.ua

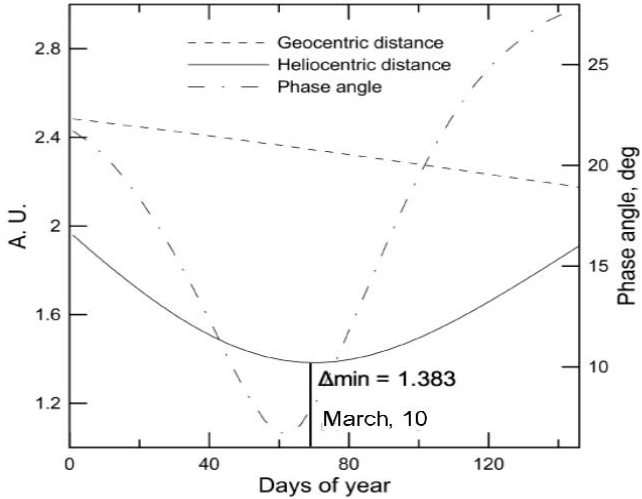

Fig. 1 Orbital circumstances of the asteroid 6478 Gault during January – May 2019.

## 2 Observations

We observed 6478 Gault using the 0.7 - m Cassegrain reflector at the Kyiv Comet Station (MPC code – 585) with FLI PL 47-10 CCD camera located in prime focus (F = 2828 mm) of the telescope and equipped by Johnson $V$ and $R$ broadband filters. The detector consists of a 1024 1024 array of 13 μm pixels, which corresponds to a scale of 0.947$''$ per pixel. Estimated readout noise is about 10 $\bar{e}$ and the conversion gain of about 1.2 $\bar{e}$/ADU.

Image processing was done with the Astroart 4.0 software (http://www.msb-astroart.com/), while stacking the frames was done using Astrometrica 4.0 (http://www.astrometrica.at/). We used ATV IDL (Barth et al. 2001) routines for aperture photometry of comets and reference stars.

The APASS (R9) star catalog was used as a photometric reference (Henden et al. 2016). This catalog includes magnitudes of stars from about 7th magnitude to about 17th magnitude in five filters: Johnson $B$ and $V$, plus Sloan $g'$, $r'$, $i'$. It has mean uncertainties of about 0.07 mag for $B$, about 0.05 mag for $V$ and less than 0.03 mag for $r'$. We used the transformation formula from $r'$ to $R_C$ magnitudes suggested by (Munari et al. 2014). About 6 – 8 reference stars of 13th – 17th magnitude were used for each night.

In Table 1 we give the values of observational dates, exposure time and number of frames for $V$ and $R$ filtered images, heliocentric distance $r$, geocentric distance $\Delta$, phase angle $\alpha$, observational apparent $V$ and $R$ magnitudes and $V – R$ colors.

## 3 Data analysis

The results of observations for several nights during January – April 2019 are showed in Table 1. We used orbital elements $r$, $\Delta$ and $\alpha$ from the Minor Planet Center database

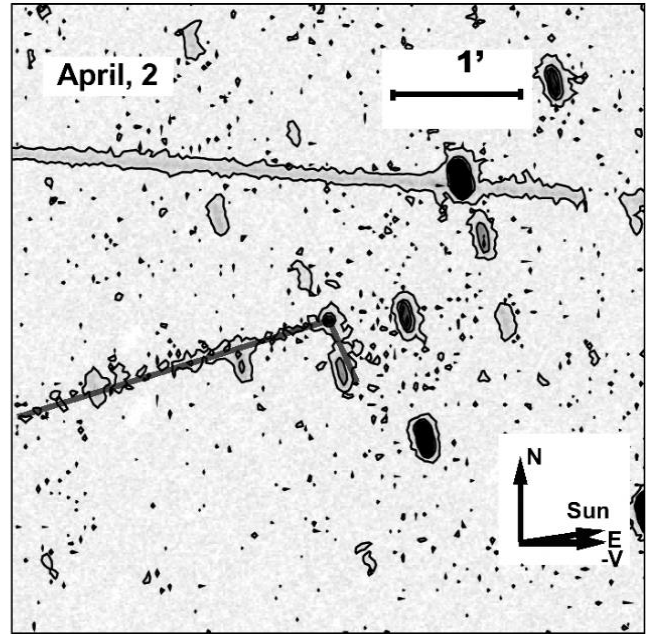

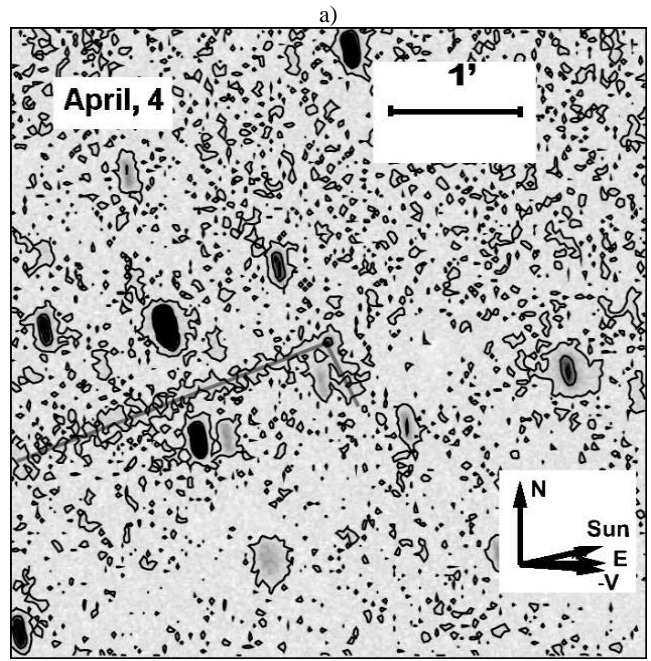

Fig. 2 Images of two-tailed asteroid 6478 Gault obtained in April, 2 (a) and April, 4 (b). Isophotes were used to enhance of visibility for the tails. The tails are marked by solid lines. Left image contains a track of the artificial satellite.

(https://minorplanetcenter.net/iau/MPEph/MPEph.html).
The maximum of brightness ($R$ = 17.17 mag) was registered close to approach with Earth (Fig. 1). Calculated color indices $V – R$ had low changes accordingly to estimations errors and were about 0.35 – 0.45 in February – April 2019. Such values are some bluer than for Jupiter-family comets or quasi-Hilda objects at perihelion (Lamy & Toth 2009; Borysenko et al. 2019).

Table 1  Log of observations and results of measurements.

| Date | $N_V \times t_{exp}$[a] | $N_R \times t_{exp}$[b] | $r$[c], A. U. | $\Delta$[d], A. U. | $\alpha$[e], deg. | $V$[f] | $R$[g] | $V - R$ |
|---|---|---|---|---|---|---|---|---|
| 2019 - 01 - 22 | – | 30 × 30 | 2.443 | 1.687 | 17.9 | – | 17.60 ± 0.06 | – |
| 2019 - 02 - 07 | 24 × 60 | 16 × 60 | 2.409 | 1.517 | 12.7 | 17.87 ± 0.02 | 17.44 ± 0.08 | 0.43 ± 0.08 |
| 2019 - 02 - 19 | – | 20 × 60 | 2.386 | 1.440 | 8.9 | – | 17.17 ± 0.05 | – |
| 2019 - 04 - 02 | 28 × 60 | 20 × 60 | 2.294 | 1.450 | 16.9 | 17.98 ± 0.02 | 17.63 ± 0.02 | 0.35 ± 0.03 |
| 2019 - 04 - 04 | 27 × 60 | 20 × 60 | 2.289 | 1.462 | 17.7 | 17.94 ± 0.02 | 17.50 ± 0.02 | 0.45 ± 0.03 |

[a] Number of stacking images and exposure of the each for V filter
[b] Number of stacking images and exposure of the each for R filter
[c] Heliocentric distance
[d] Geocentric distance
[e] Phase angle
[f] Apparent visual magnitude
[g] Apparent red magnitude

Marsset et al. 2019 obtained a temporally negative spectral slope for some near-infrared observations of the asteroid in March – April 2019. The presence of fast rotation of the object supports the hypothesis of fresh dust-free surfaces as the origin of blue color among Q-type asteroids. Q-type asteroids are spectrally more similar to ordinary chondrite meteorites than any other asteroid type. Usually, they have a strong, broad 1-micrometre olivine and pyroxene feature, and a spectral slope that indicates the presence of metal. Spectral comparisons of Gault with other blue Q-type asteroids in the SMASS (Small Main-Belt Asteroid Spectroscopic Survey) database suggested that the asteroid surface partially lost its dust regolith during the outburst events, thereby exposing a fresh blocky material at its surface (Marsset et al 2019; Prymek 2018). Our observations in April, 2 show some bluer color of the asteroid and are close to Marsset's observations date (31 March), but to confirm negative spectral slope effect by optical filters additional observations of the asteroid are needed with high SNR ratio.

The second tail of the object appeared in February. Its position gradually deviated from the direction to the main tail to about 100 degrees in early April. The direction of secondary tail does not match with direction opposite to the motion of asteroid along of its orbit. The orientation of the secondary tail can be explained by rotation of the object and may lie in the plane of rotation. The third tail was not observed due to low SNR ratios of obtained images. The low tail activity of the asteroid was detected by Toshihiko Ikemura and Hirohisa Sato (http://www.aerith.net/comet/catalog/A06478/2020-pictures.html) in early May.

Isophoted images of the 6478 Gault are presented on Fig.2. To determine the change in brightness along of the tail with distance from the core, we made aperture measurements at reference points with a constant step avoiding background stars. We used the same aperture radius as for nucleus measurements (4 arcsec), which was determined by the growth curve method (Mommert 2017). We obtained a minimum in $V - R$ color for both dates at a nucleocentric distance about 19 arcsec ( ~ 20 000 km). Slightly bluer color index (more bluer as for other points, Fig.3) in this region can be as result of the presence of some tiny gas shell around of the asteroid or some lower dust level. But obtained results are near the margin of errors. Usually, Gault did not show any evidence of gas presence by spectral or photometric observations (Jewitt et al. 2019; Lee 2019).

To explain direction and length of the long thin tail we used Finson – Probstein diagrams (Finson & Probstein 1968), which usually apply for comets, presented in celestial coordinates of the asteroid (Fig.4). The diagrams calculated for the middle exposure dates using orbital elements from *MPO* 531061. Finson & Probstein (1968) proposed a model which describes the full tail geometry with a grid of synchrones and syndynes. The lines representing respectively the locations of particles released at the same time, or with the same (Equation (1)). This model is simple because it considers only particles released in the orbital plane of the comet, and with zero initial velocity, but it provides a very good approximation of the shape of the tail, and has been used successfully to study of cometary tails (Kramer et al. 2017; Chu et al. 2020). In the tail, dust and gas are decoupled and the only significant forces affecting the grain trajectories are the solar gravity and radiation pressure $P_{radiation}$. Both forces depend on the square of the heliocentric distance but work in opposite directions. Their sum can be seen as a reduced solar gravity, and the equation of motion is simply

$$m \times a = (1 - \beta) \times g_\odot \qquad (1)$$

where $m$ is the mass of the dust particle, $a$ is the acceleration of the dust particle, $\beta$ is the ratio $P_{radiation}/g_\odot$, and is inversely proportional to the size of the grains for particles larger than 1 $\mu m$ (Vincent 2014). For each night we have a long thin tail coinciding with one of the synchronies (Fig.4).

## 4 Discussion and conclusions

As the asteroid spins faster and faster, the centrifugal force pulling the asteroid apart eventually becomes more powerful than the gravitational force holding it together. When

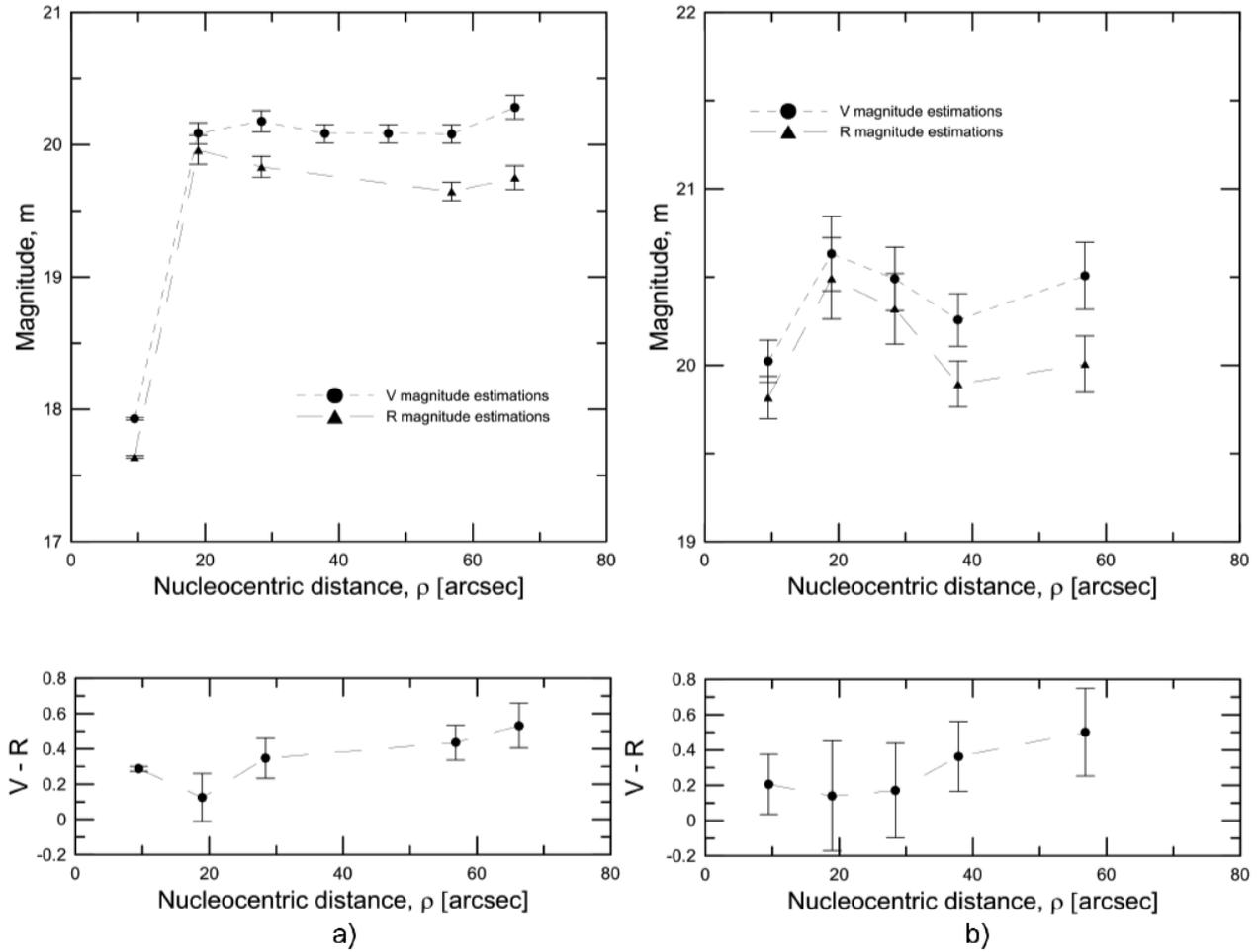

Fig. 3 Distribution of brightness along of the long tail (inner region) for April, 2 (a) and April, 4 (b).

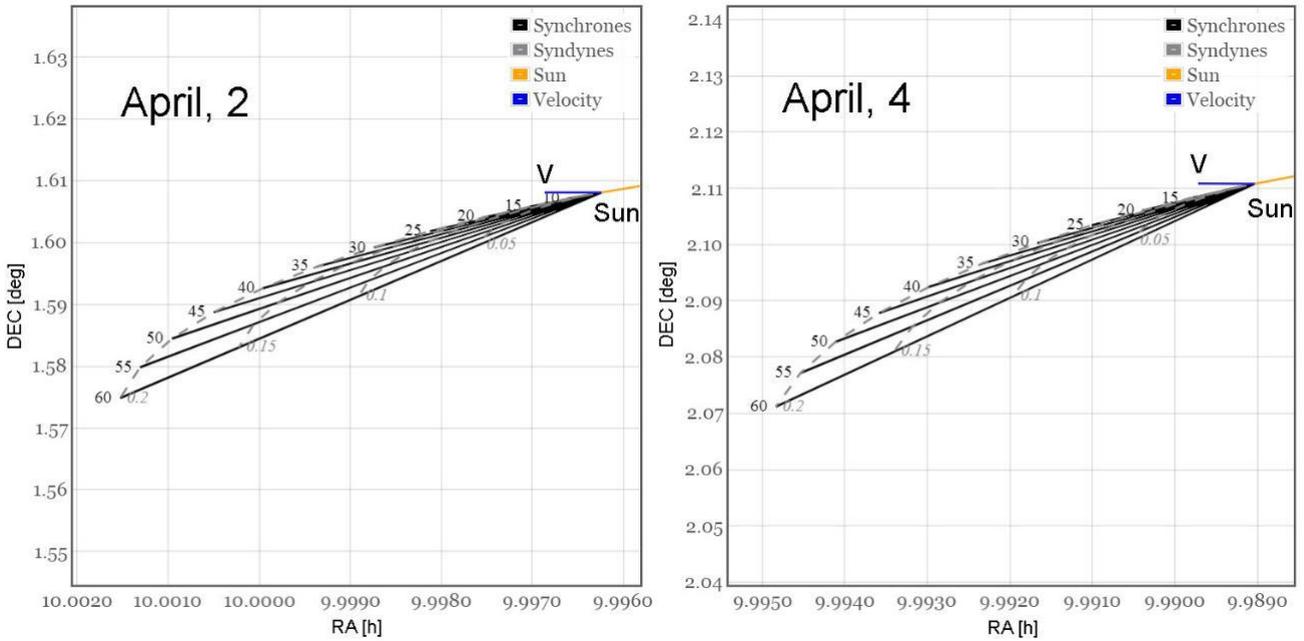

Fig. 4 Finson – Probstein diagrams for asteroid 6478 Gault. Syndynes (dashed lines) for = 0.05, 0.1, 015, 0.2 are shown. Synchrones are marked by solid lines.

Table 2 Orbital parameters of selected active asteroids.

| Date | $a^a$, A. U. | $e^b$ | $i^c$, deg. | $P^d$, years | $p^e$, hours | $T_J^f$ | Activity | References |
|---|---|---|---|---|---|---|---|---|
| 6478 Gault | 2.31 | 0.19 | 22.81 | 3.50 | 2 | 3.461 | December 2018 | Chandler et al. (2019) |
| 133P/Elst – Pizzarro | 3.16 | 0.16 | 1.39 | 5.62 | 3.5 | 3.185 | August 1996 | Hsieh et al. (2004) |
| 311P/PANSTARRS | 2.19 | 0.12 | 4.97 | 3.24 | 2 | 3.660 | September 2013 | Jewitt et al. (2018) |
| 324P/La Sagra | 3.09 | 0.15 | 21.42 | 5.44 | 3.8 | 3.100 | April 2015 | Jewitt et al. (2016) |
| 331P/Gibbs | 3.00 | 0.04 | 9.74 | 5.21 | 3.2 | 3.229 | March 2012 | Drahus et al. (2015) |
| 354P/LINEAR | 2.29 | 0.13 | 5.26 | 3.46 | 11.4 | 3.583 | February 2009 | Jewitt et al. (2010) |

$_a$ Semimajor axis
$_b$ Eccentricity
$_c$ Orbital inclination
$_d$ Orbital period
$_e$ Rotation period
$_f$ Tisserand parameter

this happens, rocks and dust are pulled of the surface of the asteroid and create a tail of debris. Multiple tails can have causes unrelated to rotational spin-up (e.g., thermal fracture exposing volatiles), but by Moreno et al. 2019 results the ice sublimation seems rather improbable. Ye et al. 2019 using of the dust dynamics code showed, that the dust grains are released just beyond the gravitational escape speed of Gault. A constant ejection speed through the particle size range is in line with the behavior of other active asteroids with non-sublimation-driven ejection (Jewitt et al. 2015).

We selected some objects with morphologically similar activity for the last more than 20 years. All of them had one or more long straight tails and period of activity about 3 – 5 months.

*133P/Elst – Pizarro* is a first discovered object with abnormally activity. The cometary nature was first discovered when a linear dust feature was observed with the ESO 1-metre Schmidt telescope at La Silla Observatory in August 1996. Around the next perihelion in November 2001, the cometary activity appeared again, and persisted for 5 months (Hsieh et al. 2004; Hsieh et al. 2010).

*311P/PANSTARRS* – main-belt comet discovered by the Pan-STARRS telescope on 27 August 2013. The multiple comet tails were observed by the Hubble Space Telescope during September 2013 – February 2014. The tails are suspected to be streams of material ejected by the asteroid as a result of a rubble pile asteroid spinning fast enough to remove material from it (Jewitt et al. 2018).

*324P/La Sagra*. The comet was originally observed to be active in 2010 – 2011. Last activity was confirmed by observations of a cometary short straight dust tail in May – June 2015 (Hsieh & Sheppard 2015; Jewitt et al. 2016).

*331P/Gibbs*. Observations in 2014 by the Keck Observatory showed that the comet was fractured into 5 pieces and rotating rapidly, with a rotation period of only 3.2 hours. Due to the YORP effect, 331P/Gibbs had begun to spin so quickly that, being a likely rubble pile, parts began to be thrown off, leaving a very long dust trail (Drahus et al. 2015).

*354P/LINEAR*. Was extremely active in January 2010. The position of the nucleus was remarkable for being off-set from the axis of the tail and outside the dust halo, a situation never before seen in a comet. The tail is created by millimeter-sized particles being pushed back by solar radiation pressure (Jewitt et al. 2010; Snodgrass et al. 2010; Yoonyoung et al. 2017). Due to the unusual morphology, this object is very unlikely to be driven by the sublimation of water ice. Several faint fragments were observed by Kim et al. (2017) in January 2017 using the 8.1-m Gemini North telescope on Mauna Kea in Hawaii.

In Table 2 we give the values of some orbital elements for these objects: semimajor axis $a$, eccentricity $e$, inclination $i$, orbital period $P$, rotation period $p$, Tisserand criterion $T_J$ calculated by the orbital elements and date of the most significant activity registered firstly in the one of the last appearances. We used orbital elements from the Minor Planet Center database (https://minorplanetcenter.net/iau/MPEph/MPEph.html).

The rotation period of the asteroid Gault is about 2 hours (Kleyna et al. 2019) and close to other active bodies (Tab.2). Some larger rotational period for Gault (3.36 hours) was obtained by Ferrin et al. 2019. So long rotation period about 11.4 hours registered for 354P/LINEAR comet give evidence of the small possibility of YORP effect influence to activity of the object. Similar activity was only once observed for active objects from the list and could be the result of partial destruction of the nucleus (354P/LINEAR, Jewitt et al. 2010). In any case, the dust activity of the objects decreased in subsequent appearances by results of observations (Hsieh et al. 2010; Hsieh & Sheppard 2015).

Gault has high inclination ($i = 22.81°$) of orbit and activity of the object was registered at high ecliptical latitude (-17 – -18 degrees in November – December 2018). Therefore, is not any possibility to connect activity of the asteroid with solar activity unlike Jupiter family comets (Musiichuk & Borysenko 2019). On the other hand, most of main-belt active asteroids have inclination less than 10 degrees (except of someone, such as 324P/La Sagra) and nearcircular orbits, therefore their activity are low connected with perihelion distance.

The results of our observations are in agreement with others (Hui et al. 2019 (apparent magnitudes and tails spatial orientation); Jewitt et al. 2019 (apparent magnitudes and colors); Ye et al. 2019 (tails spatial orientation). By indirectly factors (long time multiple tails observed and their spatial orientation) we can confirm the statement that the asteroid 6478 Gault is in rotational instability due to the Yarkovsky – O'Keefe – Radzievskii – Paddack (YORP) effect.

## 5 Acknowledgments